\documentclass[twocolumn]{aastex63}
\usepackage{CJKutf8}
\usepackage{amsmath}

\newcommand{\sgra}{$\rm Sgr~A^{\star}$}

\received{Sep 1, 2020}
\revised{Apr 2, 2021}
\accepted{Apr 6, 2021}

\submitjournal{ApJ}

\shorttitle{}
\shortauthors{Zheng, Lin, Mao}

\graphicspath{{./}{figures/}}

\begin{document}
\begin{CJK*}{UTF8}{gbsn}

\title{The influence of the secular perturbation of an intermediate-mass companion: II.  Ejection of hypervelocity stars from the Galactic Center}

\author[0000-0002-7814-9185]{Xiaochen Zheng （郑晓晨）}
\affiliation{Department of Astronomy, Tsinghua University, Beijing, China}

\author[0000-0001-5466-4628]{Douglas N. C. Lin (林潮)}
\affiliation{Department of Astronomy and Astrophysics, University of California, Santa Cruz, USA}
\affiliation{Institute for Advanced Studies, Tsinghua University, Beijing, China}
 
\author[0000-0001-8317-2788]{Shude Mao (毛淑德)}
\affiliation{Department of Astronomy, Tsinghua University, Beijing, China}
\affiliation{National Astronomical Observatories of China, Chinese Academy of Sciences,
20A Datun Road, Beijing, China}
 
\begin{abstract}
There is a population of stars with velocities in excess of 500 km s$^{-1}$ relative to the Galactic
center.   Many, perhaps most, of these hyper-velocity stars (HVSs) are B stars, similar to the disk and S stars in a nuclear cluster around a super- massive black hole (SMBH) near \sgra.  In the paper I of this series, we showed that the eccentricity of the stars emerged from a hypothetical disk around the SMBH can be rapidly excited by the secular perturbation of its intermediate-mass companion (IMC), and we suggested IRS 13E as a potential candidate for the IMC.  Here, we show that this process leads to an influx of stars on parabolic orbits to the proximity of \sgra~on a secular timescale of a few Myr.  This timescale is much shorter than the diffusion timescale into the lost cone through either the classical or the resonant relaxation.  Precession of the highly-eccentric stars' longitude of periastron, relative to that of the IMC, brings them to its proximity within a few Myr. The IMC's gravitational perturbation scatters a fraction of the stars from nearly parabolic to hyperbolic orbits, with respect to the SMBH.  Their follow-up close encounters with the SMBH induce them to escape with hyper-velocity.  This scenario is a variant of the hypothesis proposed by Hills based on the
anticipated breakup of some progenitor binary stars in the proximity of the SMBH, and its main objective is to account for the limited lifespan of the known HVSs.  We generalize our previous numerical simulations of this process with a much wider range of orbital configurations.  We demonstrate the robustness and evaluate the efficiency of this channel of HVS formation.  From these numerical simulations, we infer observable kinematic properties for the HVSs.
\end{abstract}


\section{Introduction} \label{sec:intro}

In recent years, a population of hyper-velocity stars (HVSs) has been found in the 
Galactic halo \citep{Brown2005, Brown2015, Zheng2014, Huang2017}.  Many, if not most of them are B stars with limited main sequence life span ($\lesssim 10^{1-2}$ Myr).  They are moving relative to the Galactic center with speeds in excess of 500 km s$^{-1}$.  
Their inferred travel timescale from the Galactic 
Center (GC) region to their present-day location (over a distance of $\sim 10^{1-2}$ kpc) is a 
fraction of the main sequence life span.  

Several hypotheses have been proposed for the origin of the kinematic properties of HVSs.  
The most-widely cited Hills (1988) hypothesis is based on the pre-discovery prediction of HVSs as ejected members of tidal disrupted close binary systems during their close encounters with super-massive black holes (SMBH) located at \sgra. In order to escape from its intense gravitational field, these binaries must acquire nearly parabolic orbits prior to their close encounters with the SMBH
\citep{Bromley2006}.  Such large eccentricities may be excited through classical or resonant relaxation \citep{Rauch1996} analogous to the diffusion of stars into the loss cone in the context of tidal disruption events \citep{Frank1976}.  Since these diffusion processes generally require timescales considerably longer than a few Myr,  older, 
low-mass stars may provide a greater supply of progenitors for the HVSs.  Although there is a population of low-mass main-sequence-star candidates with large radial velocities \citep{Li2015, Du2019}, 
recent GAIA data indicate that not all but most of them are likely to be bound to the Galaxy
\citep{Boubert2018}.

Some alternative scenarios for the dynamical origin of the HVSs attribute the scattering culprits to a
hypothetical intermediate-mass companion (IMC) \citep{Yu2003} or some stellar-mass black-hole
\citep{Oleary2008} companion(s) to \sgra.  Through three-body interaction, some HVSs, including surviving close binaries may be ejected \citep{Lu2007}. 
\cite{Levin2006} discussed the possibility of ejecting high-velocity stars from the GC via an inspiring intermediate-mass black hole within a few dynamical friction timescales.
Another class of hypothesis accredits the HVSs' 
formation to the evolution of very close binary stars \citep{Tauris1998}.  When the more massive primary stars undergo an off-center explosion, their less massive companions may be ejected from the proximity of SMBH due to the recoil motion \citep{Baruteau2011}.  Similar to Hill's 
hypothesis, most of these scenarios do not directly account for why the HVSs are mostly
B stars other than the fact that massive stars are mostly in binary and multiple systems \citep{Chini2012}.

In an attempt to address the origin of the hyper-velocity B stars, we pay attention to their age similarities
with a population of very young massive and Wolf-Rayet stars with estimated ages of $4-6$~Myr inside the central $\lesssim 0.5$ pc \citep{Ghez2003a}. At least half of these stars are orbiting around the central SMBH in a relatively flat clockwise disk.  These stars are dubbed as clockwise stars (CWSs) \citep{lockmann2009}. 
The possibility of other stars in an inclined counter-clockwise disk has also been suggested \citep{paumard2006} 
albeit the existence of the second disk remains controversial \citep{lu2006, lu2009}.

It has been suggested that the coeval CWSs probably formed in a common natal disk with nearly circular orbits 
\citep{Goodman2003, Levin2003}.  But, several of these stars are observed with large eccentricities ($\gtrsim 0.7$) 
\citep{lu2006, paumard2006,  gillessen2009, Do2013}.  Local dynamical relaxation processes cannot excite such large eccentricities within the short time interval of a few Myr \citep{Freitag2006, Alexander2007,  Madigan2011}.  
\cite{Yu2007} considered the dynamical perturbation of an inward-migrating intermediate-mass companion (a black hole or a dark cluster) under the influence of dynamical friction. They found that, along the perturber's 
migrating path, many stars of this stellar population could be captured into its mean motion resonance and endure the excitation 
both in eccentricity and inclination.  In order to accommodate the young age of the eccentric stars in the Galactic center,
this scenario requires a short migration time scale.  As a variant, \citet{Zheng2020} suggested in Paper I of this series 
that the IMC's secular perturbation can significantly excite the eccentricity of the young stars on a time scale comparable 
to their age, in the absence of any migration.  Since this interaction is long-range over many orbits, the IMC can be either 
a compact stellar cluster or a point mass. We speculate that the culprit is
IRS 13E which has an estimated mass $M_{\rm IMC} \simeq 10^4 M_\odot$ and is located at a projected separation 
of $\sim 0.13$~pc from the \sgra ~\citep{Krabbe1995, Genzel2003, maillard2004, Schodel2005, tsuboi2017},
albeit its integrity and mass estimates remain controversial\citep{wang2020, zhu2020}.

We show that the IMC's perturbation on the stars accumulates in time near its secular resonance, where the precession of its orbit around the SMBH matches that of the stars induced by it.  
Persistent and protected angular-momentum extraction from the stars by the IMC substantially increases their eccentricities.  During the depletion of the stars' natal disk, the location of secular resonance sweeps towards \sgra~over 
an extended region, inducing high eccentricity for a large fraction of CWSs. This effect is analogous to that induced by Jupiter on the asteroids in the solar system \citep[e.g.,][]{zheng2017a, 
zheng2017b}.  

The numerical results from Paper I also show that some disk stars become unbound to the SMBH with $e_\ast > 1$.
In this paper, we carry out a systematic study on the ejection probability of young stars formed in disks with
nearly circular orbits around \sgra.  We consider various contributions to the gravitational potential, including that
from the super-massive black hole (SMBH), IMC, the Galactic bulge, disk, and halo, as well as that from a gaseous disk out of which the disk stars emerge (\S\ref{sec:potential}).
We carry a series of numerical simulations on the secular interaction between the IMC and stars emerging
from a hypothetical gaseous disk with circular orbits.  We show the rapid excitation of their eccentricity
and the probability of close periastron passage near the SMBH in \S\ref{sec:planetaryeq}.  We present the results of several series of numerical simulations for a range of initial geometry and potential boundary
condition in \S\ref{sec:highv}.  We derive the ejection probability, the spatial distribution, and trajectories of these HVSs.  Finally in
\S\ref{sec:summary}, we summarize our results and discuss their 
implications.

\section{Numerical models}
\label{sec:potential}

A detailed description of the method and model parameters are given in the paper I of this series.  
We briefly recapitulate various components of the gravitational potential used in our calculation.
We numerically integrate the Newtonian equation of motion with the open-source \textit{N}-body code REBOUND \citep{reinliu2012} 
and its built-in MERCURIUS integrator \citep{rein2019}.  

\subsection{Prescription for the gravitational potential}

\begin{deluxetable*}{ccccccc}
\tablenum{1}
\tablecaption{Parameters for various numerical simulations \label{tab:models}}
\tablewidth{0pt}
\tablehead{
\colhead{models} 
&\colhead{${s^{\prime}_{\rm IMC} (\rm AU)}^a$} &\colhead{$i_{\rm IMC, \star} (^{\circ})^b$}  &\colhead{$i_{\star}(^{\circ})^c$}
&\colhead{${\rm Gas \ Disk \ Potential}^{d}$}  &\colhead{${\rm Gas \ Disk \ Gap}^e$}   &\colhead{{\rm Halo Potential} $^f$}   
}
\startdata
NGD         &  0    & 0  & 0   &  $-$ & $-$ & Spherical\\
fiducial    &  0    & 0 & 0   &  $+$  & $+$ &  Spherical\\
SOFT        &  5000 & 0  & 0   &  $+$ & $+$ & Spherical\\
CCW         &  0    & 180 &  0   & $+$ & $-$ & Spherical\\
CCW-NGD     &  0    & 180 & 0  &  $-$ & $-$ & Spherical\\
I30         &  0    & 30  & 0  & $+$ & $-$ & Spherical\\
I30-NGD     &  0    & 30  & 0   &  $-$ & $-$ & Spherical \\
I60         &  0    & 60  & 0  & $+$ & $-$ & Spherical\\
I60-NGD     &  0    & 60  & 0  &  $-$ & $-$ & Spherical\\
I60-IS60    &  0    &  60 &  60   & $+$ & $-$  & Spherical\\
I60-TRI     &  0    &  60 &  0   & $+$  & $-$ & Triaxial\\
I120        &  0    & 120  & 0  & $+$ & $-$ & Spherical\\
I120-NGD    &  0    & 120  & 0   &  $-$ & $-$ & Spherical\\
\enddata
\tablecomments{ $^a$ The softening parameter of IMC's potential. $^b$ The initial mutual inclination between the orbits of IMC and disk stars. $^c$ The initial inclination of the orbit of disk stars (gas disk) relative to the Galactic disk (reference plane).$^d$ The presence of the gas disk potential.$^e$ The presence of a gas-free gap around the IMC's orbit. $^f$ The form of the halo potential. }
\end{deluxetable*}

For different components of the Galactic potential, we adopt the conventional prescription for the SMBH, IMC, Galactic bulge, disk, and halo such that
\begin{equation}
\begin{split}
\Phi_{\rm SMBH} & = - \frac{G M_{\rm SMBH}}{r} + {\rm post \ Newtonian \ terms}, \\
 \Phi_{\rm IMC} & = - \frac{G M_{\rm IMC}}{\sqrt{\mid \vec{r}-\vec{r}_{\rm IMC} \mid^2 + s^{\prime 2}_{\rm IMC}}},\\ 
 \Phi_{\rm bulge} & = - \frac{G M_{\rm bulge}}{r+a_{\rm bulge}}, \\
 \Phi_{\rm disk} & = - \frac{G M_{\rm disk}}{\sqrt{R^2 + \left(a_{\rm disk}+\sqrt{z^2 + d_{\rm disk}^2} \right)^2}}, \\ 
 \Phi_{\rm halo} & = - \frac{G M_{\rm halo}}{r} {\rm ln} \left(1 + \frac{r}{a_{\rm halo}} \right).
 \label {eq:galactic_pot}
\end{split}
\end{equation}
In the above expression, $r$, $R$, and $z$ are the spherical radius, cylindrical radius 
and the distance above the disk; $r_{\rm IMC}$ is the distance of the IMC from the GC.
In Table 1 of Paper I, we specify $a_{\rm bulge} (=0.6$ kpc), 
$a_{\rm disk} (=5$ kpc), $d_{\rm disk} (=0.3$ kpc), and $a_{\rm halo} (=20$ kpc) are 
the scaling lengths for the bulge, disk, and halo
respectively; $M_{\rm SMBH} (=4 \times 10^{6} M_\odot)$ and $M_{\rm IMC} (=10^4 M_\odot)$ are the mass of the 
SMBH and IMC, $M_{\rm bulge} (=10^{10} M_\odot)$, $M_{\rm disk} (4 \times 10^{10} M_\odot)$, and $M_{\rm halo} (=10^{12} M_\odot)$ 
are the mass scaling factor for the bulge, disk, and halo, respectively \citep{Miyamoto1975, Hernquist1990, Navarro1997}. 
For the central SMBH, we include the post-Newtonian corrections (1$^{\rm st}$ and 2$^{\rm nd}$ orders) near its Schwarzchild radius, same as the formulae in \cite{kidder1995}. As it makes negligible effects on most orbits, the $2^{\rm nd}$-pN is only considered for those highly eccentric orbits, where the closest distance to the central SMBH is smaller than $\sim 10^{-5}$~pc, to avoid large energy changes during close encounters with the SMBH as discussed in  \cite{Rodriguez2018}.

In one of the models (SOFT), we consider the possibility of IMC being a dense cluster with a Plummer potential and a softening parameter $s^{\prime}_{\rm IMC} (=5 \times 10^3$ AU).  In all models, IMC's initial semi-major axis and eccentricity are $a_{\rm IMC} =0.35$ pc and $e_{\rm IMC}=0.3$ respectively.

We neglect the departure from spherical symmetry for the Galactic bulge.  Although a tri-axial mass distribution in the Galactic bulge can lead to IMC's precession which modifies the secular evolution of single and binary stars' orbits 
\citep{Petrovich2017,  Mathew2020}, it is negligible in comparison with the IMC's secular perturbation on the
relatively massive main-sequence stars within the central parsec from the Galactic center during their lifespan 
(i.e., over a few to tens Myr).  However, a non-spherical component in the mass distribution in the Galactic halo can significantly modify the trajectories of HVSs on a dynamical timescale.  We consider this possibility in model I60-TRI
with a triaxial potential \citep{YuMadau2007MNRAS} such that
\begin{equation}
\begin{split}
\Phi_{\rm halo} &= - \frac{G M_{\rm halo}}{r^{\prime}} {\rm ln} \left(1 + \frac{r^{\prime}}{a_{\rm halo}} \right) ,   \\
{\rm with} \ \ \ r^{\prime} &= p_{\rm halo}^{1/3} q_{\rm halo}^{1/3} \left( x^2 + \frac{y^2}{p_{\rm halo}^2} + \frac{z^2}{q_{\rm halo}^2} \right)^{1/2} , 
\end{split}
\label{eq:triaxial}
\end{equation}
where $p_{\rm halo}$ and $q_{\rm halo}$ describe the asphericity of the halo potential. If the $p_{\rm halo} = q_{\rm halo} = 1$, the halo potential is spherical. In I60-TRI model, we set $p_{\rm halo} = 0.8$ and $q_{\rm halo} = 0.7$, with z axis normal
to the Galactic disk, the same as the assumption in \cite{YuMadau2007MNRAS}.

\subsection{Gaseous and stellar disks}

For the gaseous disk, we adopt a prescription similar to our previous models of protostellar disks.
In the absence of a gap, the precession of the embedded stars is driven by the gaseous disks' central
gravity
\begin{equation}
f_{\rm emb, gas} = - 4\pi G \Sigma Z_{k} ,
\label{eq:f_gas_star}
\end{equation}
where $Z_k = 1.094$ is the same as in \cite{ward1981, nagasawa2005, zheng2017a}.  This force is also
applied to an IMC with an inclined orbit or a counter orbiting IMC. 
For a co-orbiting IMC with a semi-major axis $a_{\rm IMC} (=0.3-0.4$ pc) and eccentricity $e_{\rm IMC} (=0.2-0.4)$, we assume it opens up a gap in the disk with a width 
\begin{equation}
a_{\rm in/out} = a_{\rm IMC} (1 \mp e_{\rm IMC}) \left[ 1 \mp \left( \frac{M_{\rm IMC}}{3M_{\rm SMBH}}\right)^{1/3} \right].
\label{eq:ainout}
\end{equation}
Such that the central gravity from the gaseous disk on it becomes
\begin{equation}
\begin{split}
& f_{\rm gap, gas}   =  2\pi G \Sigma \sum_{l=0}^{\infty} \left[ \frac{(2l)!}{2^{2l} (l!)^2} \right]^2 
\left(\frac{2}{4l+1}\right)  \\
& \times \left[ (2l) \left( \frac{r}{a_{\rm out}}  \right)^{2l+1/2}  - (2l+1) \left( \frac{a_{\rm in}}{r} \right)^{2l+1/2}  \right] .
\end{split} 
\label{eq:f_gas_IMC}
\end{equation}
For illustrative purpose, we assume a power-law surface density ($\Sigma$) distribution constructed for the minimum mass solar nebula
\citep{Hayashi1985} such that
\begin{equation}
    \Sigma = \Sigma_0 \left(\frac{r}{R_0} \right)^{-3/2} e^{-t/\tau_{\rm dep}} ~\rm g/cm^2,
\end{equation}
where $\tau_{\rm dep} (=3.16$ Myr) is the depletion timescale of the gas nebula, 
$\Sigma_0 (= 600$~g cm$^{-2}$) and $R_0 (=10^3~\rm AU)$ are the fiducial surface density
and radii. 

We assume the stars emerge from the gaseous disk and the gaseous disk preserves its spin orientation.
In one model (I60-IS60), we assume the gaseous disk is inclined to the Galactic disk by $60^{\circ}$.  In all
other models, the gaseous and Galactic disks are assumed to be in the same plane.
For models where the orbital planes of the stars, IMC, and/or the Galaxy are inclined, the inclination
of the stars relative to the gaseous disk ($i_\ast$)  evolves with time.  The force in the ($z$) direction 
normal to the gaseous-disk plane is approximated with 
\begin{equation}
    f_{\rm z, gas}  = - 2\pi G \Sigma(R) \frac{ {\rm sign}(z) (z/H)^2}{1 + (z/H)^2} 
\end{equation}
where the gaseous disk's aspect ratio $H/R = (H_0/R_0) (R/R_0)^{1/4}$ with $H_0/R_0=10^{-2}$.

\begin{figure*}[ht!]
\epsscale{0.6}
\plotone{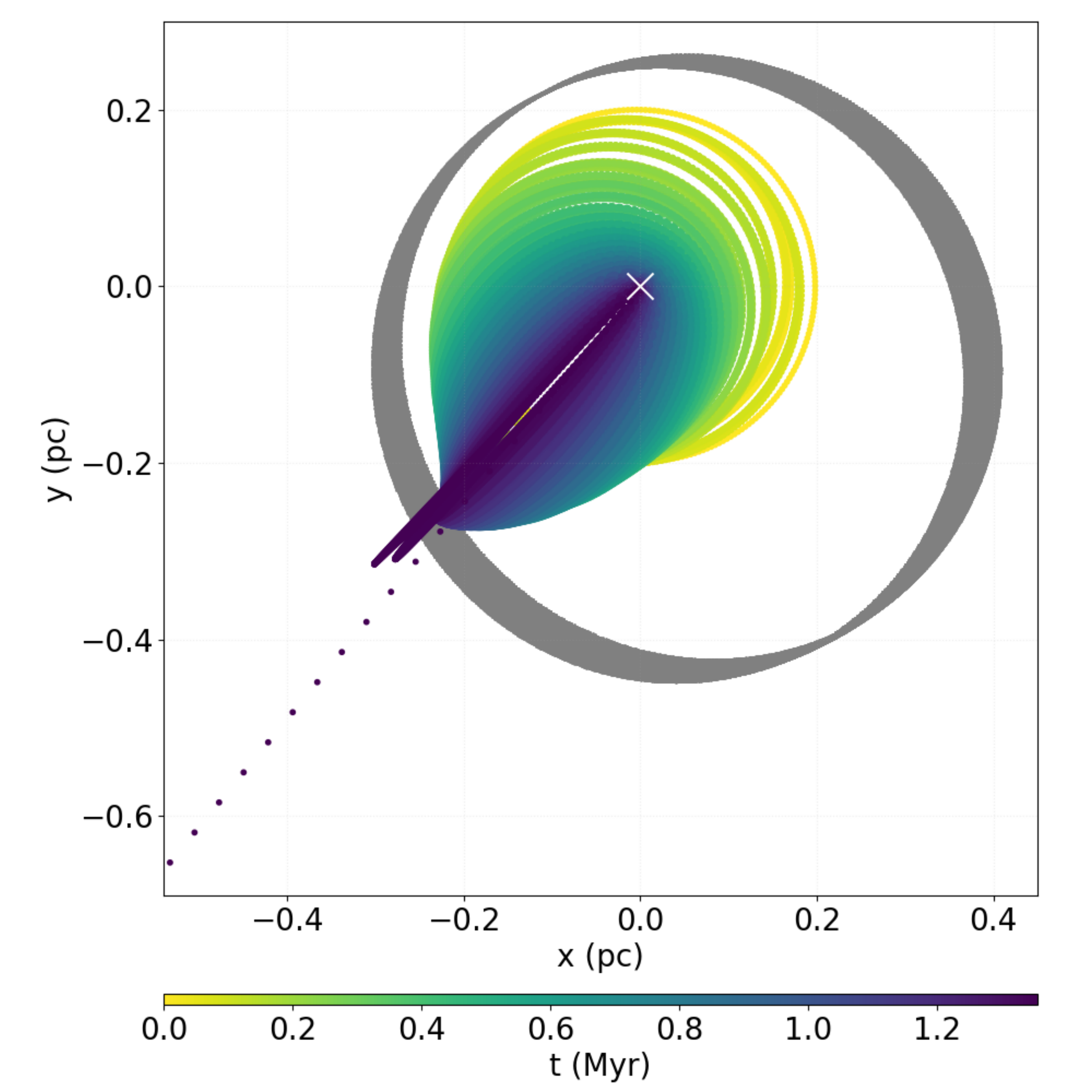}
\caption{The orbit evolution of a potential HVS. Color represents the time evolution. The white cross refers to the center of our Galaxy, the grey dots label the orbit of an IMC, with eccentricity 0.3.
\label{fig:orbit_hvs}}
\end{figure*}

\subsection{Numerical models}
In order to investigate the effects of sweeping secular resonance on the evolution of GC stars and its parameter dependence, 
we test a series of 
numerical models with various setups. Table \ref{tab:models} lists all those models, including simulations with three different gas-disk
depletion timescales.  In all cases, the stars are assumed to reside in the gaseous disk with circular orbits initially.  

We assess the influence of the sweeping secular resonance by turning off the gravity due to the gaseous disk in a comparative model NGD.  We also simulated a SOFT model in which the IMC's gravity is approximated by a Plummer potential.  
In the fiducial, NGD, and SOFT models, we assume the \sgra, companion, stars and gaseous disk are all in the same plane, and their orbital angular momentum vectors are aligned. In the CCW model, we consider the possibility a counter
orbiting IMC as inferred from some observations of IRS 13E \citep{Genzel2003}.
In a series of 3-d cases, we assume 30$^{\circ}$, 60$^{\circ}$, and 120$^{\circ}$ misalignment angle between the IMC's and the star's 
initial orbital plane.  

\section{The secular evolution of a representative star}
\label{sec:planetaryeq}
In Paper I, we show that IMC's secular perturbation rapidly and robustly leads to the eccentricity
excitation of the disk stars under a wide range of kinematic configuration.  However, to
the lowest order in $e_\ast$ and $e_{\rm IMC}$, secular interaction leads to angular momentum exchanges
with negligible energy exchange.  Under this condition, the stars' $e_\ast$ is excited without significant changes
in $a_\ast$, and the perturbed stars remain bound to the SMBH.  However, as $e_\ast$ approaches to unity,
higher-order perturbation through mean-motion resonances (MMRs) intensify. MMRs induce energy exchange
between the stars and the IMC \citep{murray2000}.  

It is also possible for stars' initial $a_\ast= a_0 \gtrsim (1-e_{\rm IMC}) a_{\rm IMC} /2$ to cross
the IMC's orbit as their $e_\ast$ approaches to unity.  The stars' longitude of periastron
($\varpi_\ast$) precess mainly under the influence of IMC's secular perturbation, the gravity due to the
Galactic bulge and a hypothetical gaseous disk, a rate 
\begin{equation}
    \frac{d \varpi_{\star}}{d t} =  \frac{d \varpi_{\star}}{d t}^{(\rm sec)} + \frac{d \varpi_{\star}}{d t}^{(\rm bulge)}  + \frac{d \varpi_{\star}}{d t}^{(\rm emb, gas)}.
\end{equation}
In the limit of small $e_\ast$, the magnitude and sign of each term on the right hand side can be 
determined in terms of the model parameters (see Paper I).  At the instant of orbit crossing (when 
the IMC is located at $\theta_{\rm IMC}$ and $r_\ast = r_{\rm IMC}$), the stars' azimuthal position 
is $\theta_\ast \simeq \pi \pm \varpi_\ast$.  An close encounter between the star and IMC would 
occur if $\vert \theta_\ast - \theta_{\rm IMC} \vert < < 1$ or equivalently $\vert \varpi_\ast-
\theta_{\rm IMC} \vert \simeq \pi$.  The strong interaction during their close encounters (with 
minimum distance between the stars and the IMC, $s_\ast < < a_\ast$) can also lead to an energy exchange 
between the stars and the IMC \citep{Lu2007}.  

The energy exchange between the IMC and the stars in its MMRs or endure close encounters with it 
leads to modest changes in their $e_\ast$ and periastron distance relative to the SMBH ($r_{\rm min}$).
They become unbound from the SMBH when their $e_\ast \geq 1$.  We illustrate this 
possibility with a representative star with an initial $a_\ast=a_0=0.2$ pc in the fiducial model 
(Table \ref{tab:models}), the orbital evolution is shown in Figure~\ref{fig:orbit_hvs}. 

The color changes from yellow to dark in Figure~\ref{fig:orbit_hvs}. It delineates the time evolution. For this representative star, 
its initial orbit is circular. Its eccentricity is gradually excited by the IMC's secular resonance, as we suggest above. In the last few orbits shown here, its eccentricity approaches to unity and its apoastron distance is outside the IMC's periastron distance.
Differential precession brings the star to the IMC's azimuthal proximity where they undergo close encounter. The energy exchange
between the star and the IMC liberates the star from SMBH's gravitational confinement.  The star's escape speed is determined by
its modified impact parameter relative to SMBH after its close encounter with the IMC.  Since the star ensures strong gravitation force
from the SMBH, we carry out a convergence test to ensure numerical accuracy.

\section{Ejection of high velocity stars}
\label{sec:highv}

The possibility of launching hyper-velocity stars (HVSs) arises from the rapid eccentricity excitation of disk stars within 1 pc via the SSR mechanism. After a fraction of stars attain nearly parabolic orbits, their periastron distance $r_{\rm min}$ is significantly reduced. Through the \citet{Hills1988} mechanism, any compact binary stellar systems may break up and toss out HVSs.  Besides, IMC's nonlinear secular perturbation may also lead to the ejection of single stars to become HVSs. Since the Hills mechanism has already been
extensively studied \citep{Kenyon2008}, we focus our investigation in this paper on the IMC-induced ejection of single stars. 

\begin{figure*}
\epsscale{1}
\plotone{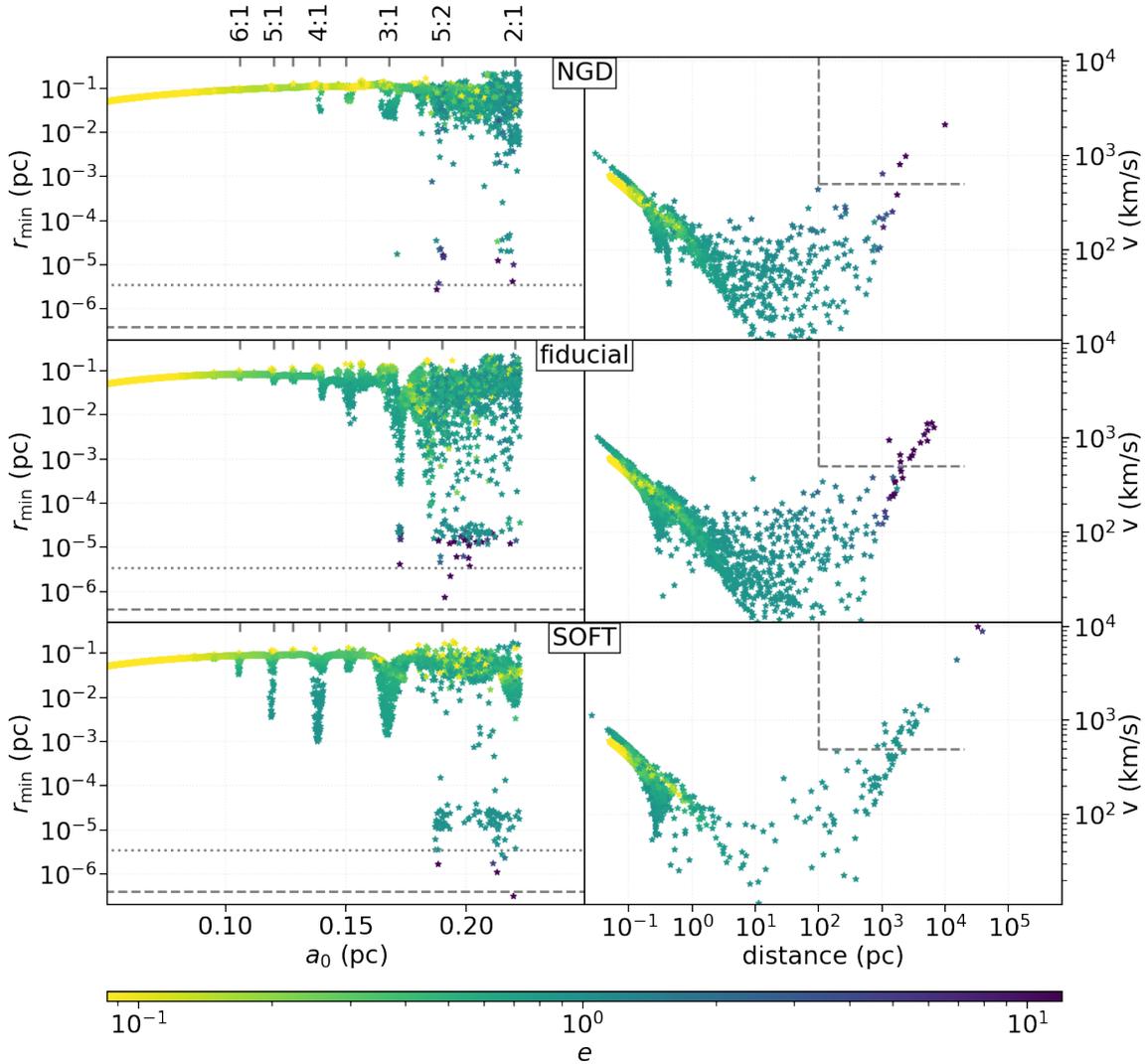}
\caption{{\bf Left panels:} The closest distance ($r_{\rm min}$) of 5000 disk stars to the \sgra~ as a function of their initial location ($a_0$). The gray dotted lines represent the tidal disruption limit for a main-sequence star with mass $12~M_\odot$, and the gray dashed lines label the Schwarzchild radius the SMBH. {\bf Right panels:} The magnitude of the total velocity of disk stars as a function of distance from the SMBH around 6 Myr. Colors label the value of orbital eccentricity at the end of the simulation: the yellow colors refer to the low-eccentricity excitation, while dark-blue colors represent hyperbolic orbits, and green colors indicate the simulated stars are excited to nearly parabolic orbit. Several critical mean motion resonance locations to the IMC are labels at the left panels. The grey dashed lines at the right panels define the criterion of an HVS, which must satisfy $v > 500~\rm km/s$ and $a > 100$~pc. 
\label{fig:a0_rmin}}
\end{figure*} 

\subsection{The influence of sweeping secular resonance}
\label{sec:influencessr}
To examine this possibility, we consider several models, which are variations of the fiducial model (Figure~ \ref{fig:a0_rmin}).
In all these models, we use $\Sigma_0=600$ g cm$^{-2}$, $\tau_{\rm dep} = 3.16$~Myr, $M_{\rm IMC}=10^4 M_\odot$, $a_{\rm IMC} = 0.35$ pc
and $e_{\rm IMC}=0.3$.  For each model, we randomly place a population of $5000$ disk stars with an initial semi-major axis $a_0$ randomly distributed between
0.05~pc and $a_{\rm in} (\sim 0.22$ pc) (Equation \ref{eq:ainout}).  We first analyze the results of the NGD model which neglect the effect of 
the sweeping secular resonance.  In this case, large eccentricity excitation occurs only for stars with orbits which are initially 
around 
2:1 and 5:2 mean motion resonances with that of the IMC (at $\sim 0.22$ and $\sim 0.19$~pc respectively).  Very few stars are ejected into 
hyperbolic orbits with asymptotic speed $v > 500$ km s$^{-1}$, after their $r_{\rm min}$ is reduced to $\sim 1$ AU (top panel in Figure~\ref{fig:a0_rmin}).

\begin{figure*}
\epsscale{0.8}
\plotone{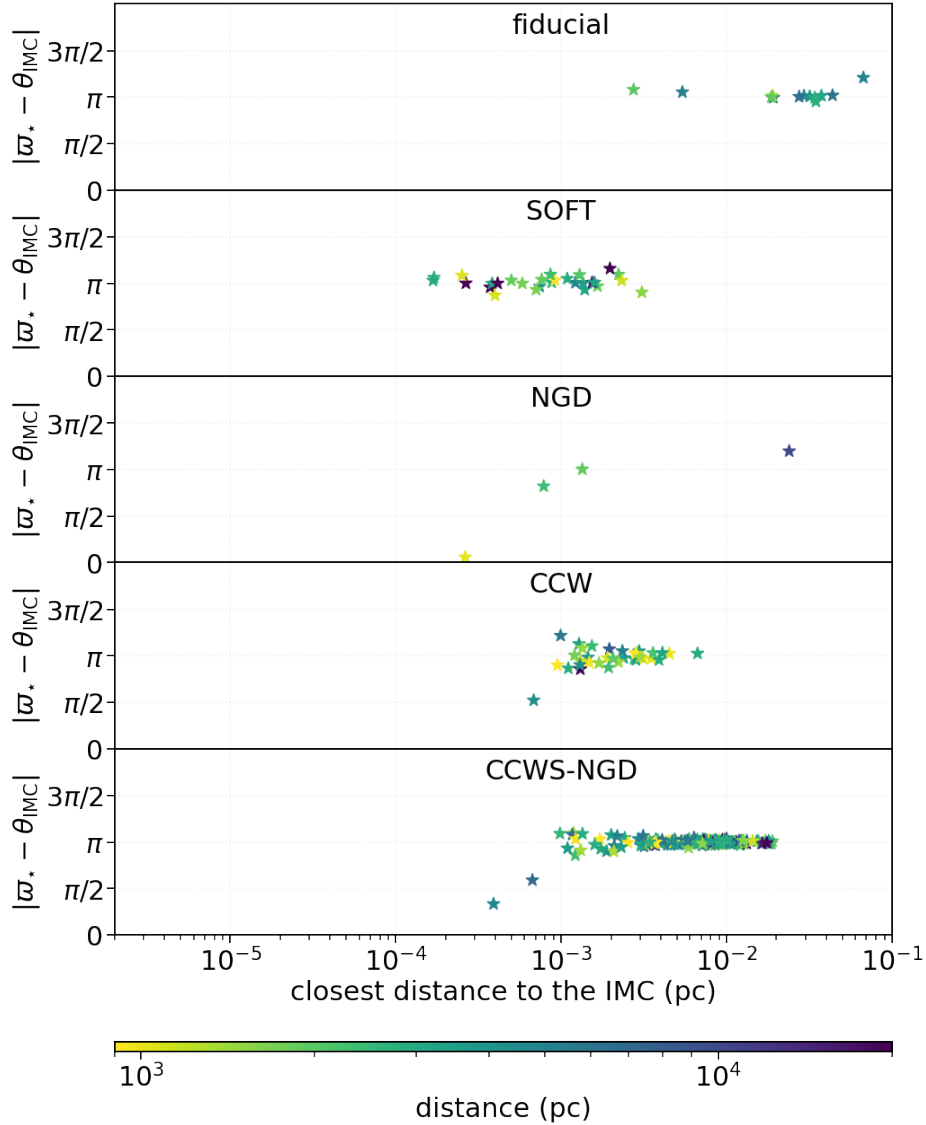}
\caption{Close encounters immediately prior to ejection.  Along the x axis, HVSs' closest distance from the IMC 
$r_{\rm IMC, \ast} = \vert {\bf r}_{\rm IMC} - {\bf r} \vert$ is measured in units of pc.  The y axis is their
$\Delta \theta_{\rm \ast, IMC}= \theta_{\rm IMC} - \omega_\ast$ immediately after the instant of their closest 
encounters (for stars on parabolic orbits, $\pi$ corresponds to colocation with the IMC).  The ejected 
HVSs' distances from \sgra~ at the end of the calculation are indicated by color.  
\label{fig:d_imc}}
\end{figure*}

In the fiducial model (middle panel in Figure~\ref{fig:a0_rmin}), the evolving gaseous disk introduces the effect of sweeping secular resonance. After 6 Myr ($\sim 2~\tau_{\rm dep}$), a population of disk stars in the region $0.15-0.22$~pc from the SMBH attains 
hyperbolic orbits ($e_\star >1$) and reaches distance $\gtrsim 100$~pc with asymptotic speed with a magnitude $v > 500$ km s$^{-1}$
mostly in the direction away from the SMBH. We refer these stars as the candidate HVSs.  We have not included, in our models, compact binary systems which are the progenitors of HVSs under the 
Hills hypothesis.  This ejection channel for HVSs through sweeping secular resonance occurs rapidly during the epoch of disk depletion.  It can account for HVSs' young ages and high masses.  Moreover, a minute fraction of stars attain $r_{\rm min}$ 
comparable to the tidal disruption radius \citep{Frank1976}, $r_{\rm tide} \simeq (M_{\rm SMBH}/m_\star)^{1/3} s_\ast 
(\approx 3.5 \times 10^{-6}$ pc). For all those very close approaching events, we re-check them with higher accuracy. These rare close encounters raise the possibility of infrequent tidal disruption events.   
 
The HVSs' asymptotic speed in the fiducial model is $\sim 1 \% $ of the escape speed for hypothetical stars on circular
orbits at $r_{\rm min}$.  However, it is slightly larger than IMC's orbital speed at IMC's periastron.  This ``slingshot'' 
is accomplished in three steps (see discussions in \S\ref{sec:planetaryeq} and illustrative Figure \ref{fig:orbit_hvs}).  
1) Most of the ``heavy lifting'' is accomplished
through IMC's secular perturbation which induces some stars to attain nearly parabolic orbits (with very little orbital
energy).  2) These marginally bound stars receive some impulsive ``gravitational assist'' from the IMC during their close encounters.  Although the fractional change on the stellar orbits ($\sim (M_{\rm IMC} a_{\rm IMC} /
M_{\rm SMBH} r_{\rm IMC, \ast})^{1/2}$ where $r_{\rm IMC, \ast} = \vert {\bf r}_{\rm IMC} - {\bf r} \vert$ is the stars' 
distance from the IMC) is modest, the IMC and SMBH's gravity together may significantly modify some stars'
energy and impact parameter relative to the SMBH.  3) Some of the scattered stars undergo an additional close encounter
with the SMBH and escape from its gravitational confinement with $e_\star > > 1$.  

\begin{figure*}
\epsscale{1}
\plotone{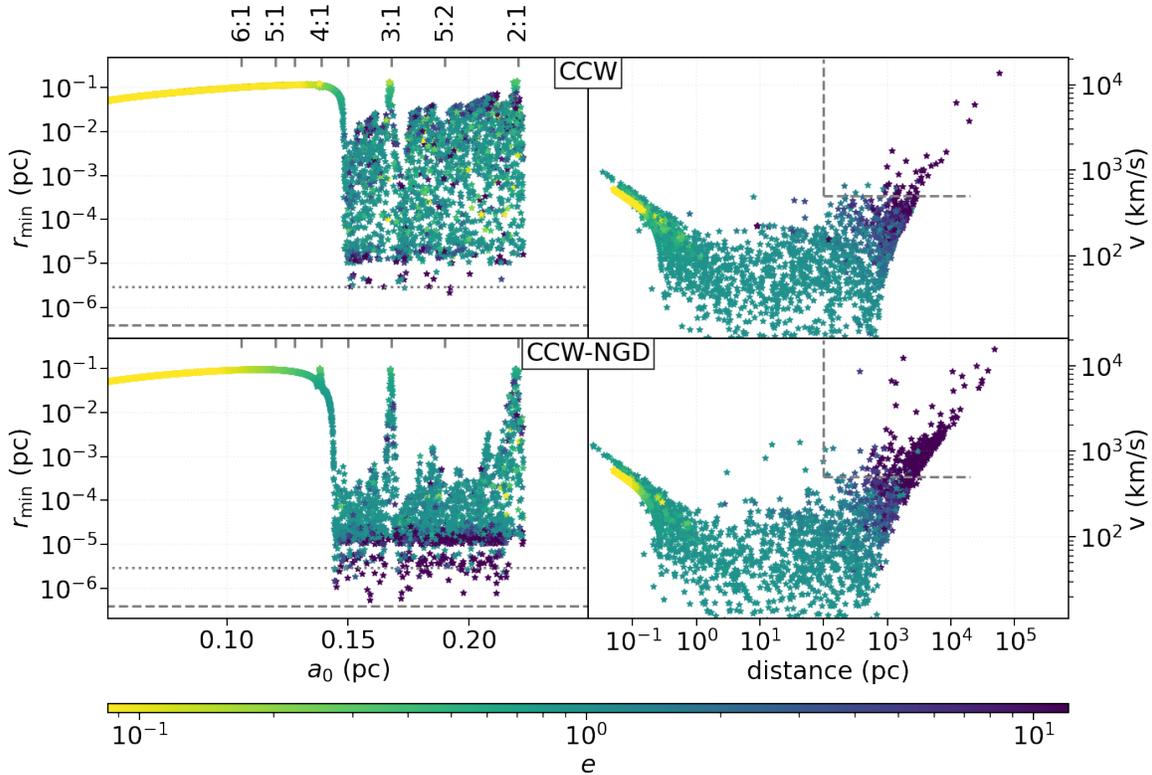}
\caption{Same as Figure~\ref{fig:a0_rmin}, with a comparison between 5000 disk stars in the CCW and the CCW-NGD models. 
\label{fig:a0_rmin_ccw}}
\end{figure*}

We select a group of HVSs and plot their distance $r_{\rm IMC, \ast}$ and angular separation $\Delta \theta_{\rm \ast, IMC}
= \vert \theta_{\rm IMC} - \omega_\ast \vert$ from the IMC during or immediately after the instant of their closest 
encounters (Figure \ref{fig:d_imc}).  
Dispersion of $\Delta \theta_{\rm \ast, IMC}$ from $\pi$ is due to the scattering during the closest encounters 
and it can attain non-negligible values for stars with orbits marginally overlapping that of the IMC.  
The ejected HVSs' distances from \sgra~at the end of the calculation are indicated by color.  The lack of
correlation between the minimum star-IMC distance and HVSs' distances from \sgra~is an indication that the IMC is 
not the primary culprit of hyper-velocity ejection.  In contrast to the correlation between $r_{\rm min}$ and the 
HVSs' final distances at the end of the simulation (Figures \ref{fig:a0_rmin} and \ref{fig:a0_rmin_ccw}) provides supporting evidences to suggest that
their scattering by the SMBH after their impulsive close encounters with the IMC was the event that led to 
their escape with hyper velocities.

In model SOFT, we adopt a Plummer model for the companion's gravitational potential.  The softening parameter
$s_{\rm IMC}^\prime \sim 0.07~ a_{\rm IMC}$ is marginally smaller than its Hills radius $r_{\rm H} = (M_{\rm IMC} / 3 
M_{\rm SMBH})^{1/3} a_{\rm IMC} \sim 0.09~a_{\rm IMC}$. 
This modified potential weakens the impact of star-IMC close encounters without sacrificing the IMC's distant secular perturbation on the stars.  In addition, we turn off the merging possibility between the IMC and disk stars only in this SOFT model. 
The bottom panel of Figure (\ref{fig:a0_rmin}) indicates that $e_\ast$ for a large fraction of stars is excited to $\sim 1$.
With their nearly zero total orbital energy, close encounters with IMC's relatively shallow potential are adequate to perturb these stars from parabolic to hyperbolic orbits.  The result of the SOFT model suggests that the IMC needs not to be an IMBH, and the distributed potential of a compact stellar cluster is a sufficient agitator for the HVSs.  
 
\subsection{Counter orbiting IMC}
We also simulate a CCW model with the identical gas and stellar disks as the fiducial model. The IMC's $M_{\rm IMC}$, $a_{\rm IMC}$, 
and $e_{\rm IMC}$ are also identical except its orbit is in the counter-clockwise direction (i.e., its orbital angular momentum is anti-parallel to that of the stars and gas disk).  A wealth of disk stars in more extended regions are significantly excited
(with $e_\star > > 1)$ by this counter-orbiting IMC, albeit in the 2-D limit. However, those disk stars initially distributed within $\sim 0.13$~pc are less perturbed compared to the fiducial model and SOFT model in Figure~\ref{fig:a0_rmin}). Moreover, the mean motion resonances are less pronounced in this model, as shown in Figure~\ref{fig:a0_rmin_ccw}. The CCW model can produce many more ejections and HVSs than the fiducial/SOFT models. The frequency of HVSs is close to $1\%$ among all disk stars (shown in Figure~\ref{fig:model_frq}).  
This result is consistent with the analytic calculation in the paper I of this series, where we showed that
a counter-orbiting IMC could indeed induce secular resonances over a wide region.  Although the presence of gas disk introduces
additional precession which initially reduces the effectiveness of IMC's secular perturbation, the influence of a counter
orbiting IMC is reinstalled after the disk depletion.  In another series of CCW-NGD model, a counter orbiting
IMC is imposed without the gas disk potential (i.e., with $\Sigma_0=0$).  The results in Figure~\ref{fig:a0_rmin_ccw} show that it
is just as capable to excite $e_\star$ and eject HVS's.

\begin{figure*}
\epsscale{1}
\plotone{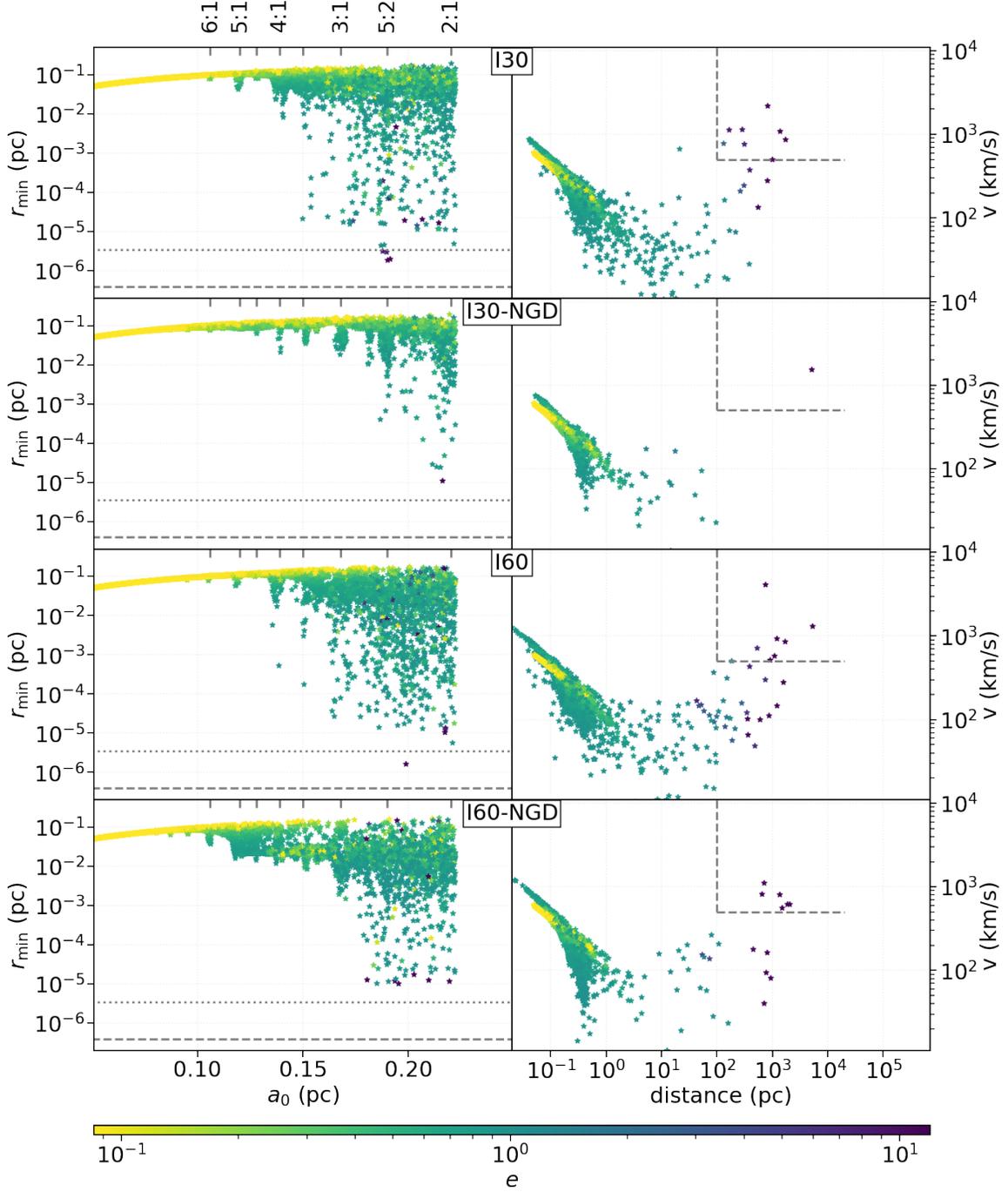}
\caption{Same as Figure~\ref{fig:a0_rmin} with a comparison between the I30, I30-NGD, I60, and I60-NGD models. 
\label{fig:a0_rmin_inc}}
\end{figure*}

\subsection{IMC on an inclined orbit}

\begin{figure*}
\epsscale{1.}
\plotone{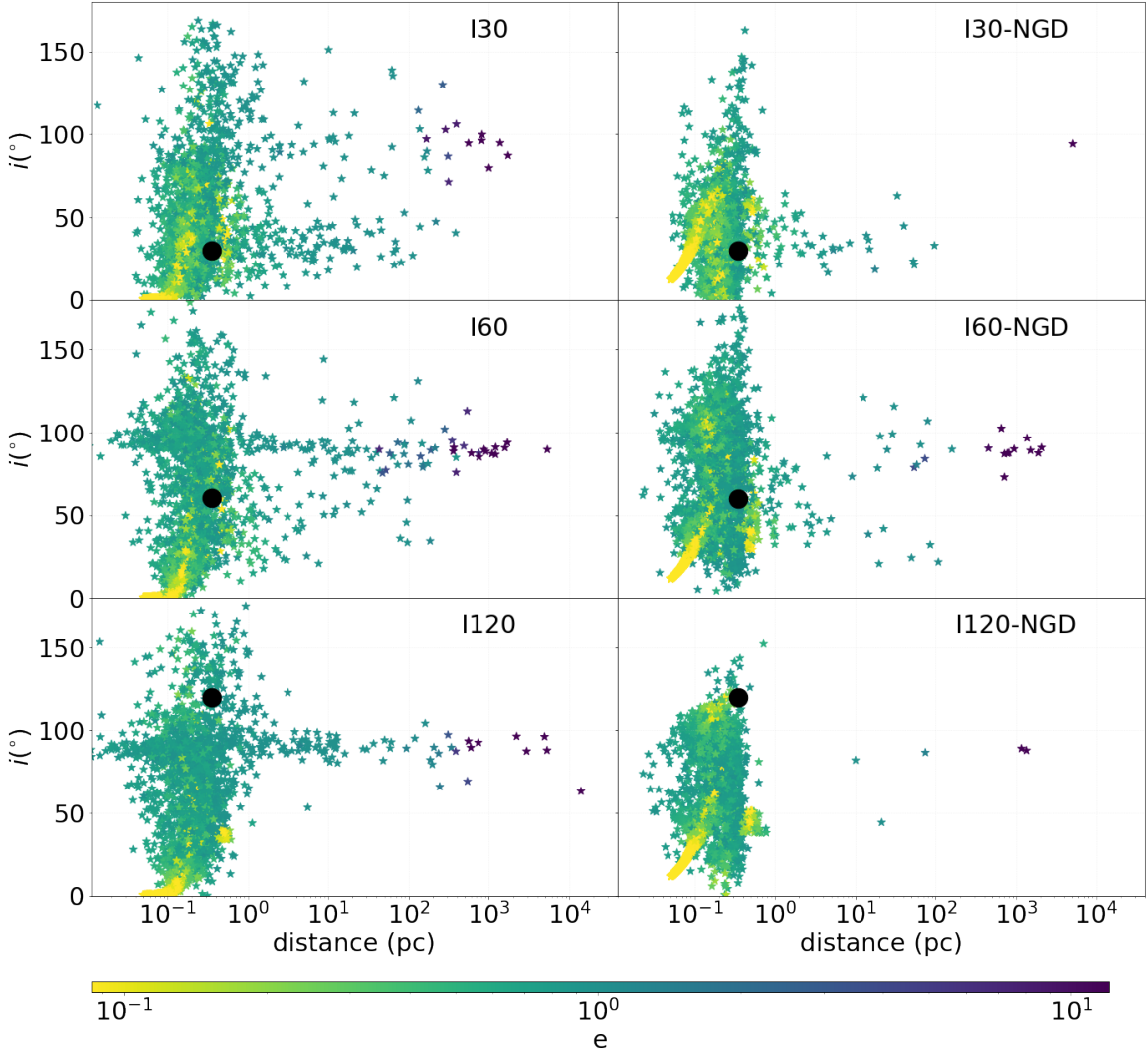}
\caption{The $(e, i)$ distribution of simulated stars around 6 Myr as a function of distance to the GC in the I30, I30-NGD, I60, I60-NGD, I120, and I120-NGD models.   Colors label the eccentricity of disk stars, and $i$ refers to the inclination of the orbits of disk stars and IMC relative to the Galactic disk plane (reference plane), range from $0^{\circ}$ to $180^{\circ}$, where $90^{\circ}$ means that the orbits of disk stars are perpendicular to the Galactic disk. The black dots represent the IMC's final semi-major axis and inclination (relative to the reference plane). 
\label{fig:d_i_e}}
\end{figure*}

In most of the above models,  we assume the simulated disk stars formed in the mid-plane of a gas disk, and they are coplanar to a co-orbiting or a counter orbiting IMC. Due to the uncertainties in the observed orbit of IRS 13E, a more general analysis is warranted.  Although we can generalize the analytic calculation in the paper I, many degrees of freedom expand its complexity.  Instead, we adopt additional illustrative series of numerical simulations. 
For models with evolving gas disks, we use the same value of $\Sigma$ distribution (but without a gap), $\Sigma_0$, and $\tau_{\rm dep}$ as the fiducial model.

In the I30, the orbits of all the disk stars are initially set to be in the same plane with $i_\ast = 0^\circ$. 
We also adopt a mutual initial inclination of $ i_{\rm IMC} - i_{\star} = 30^{\circ}$ for IMC's orbit . 
The IMC's mass, semi-major axes, and eccentricity ($M_{\rm IMC}$, $a_{\rm IMC}$, and $e_{\rm IMC}$) in this and other models with inclined stellar orbits are identical to those in the fiducial model.
Similarly, the model I60 and I120 are set with the initial mutual inclination between disk stars and the IMC, to be $60^{\circ}$ and $120^{\circ}$, respectively.
For all these inclined models, disk stars are initially embedding in the gaseous disk and coplanar to the Galactic disk (refers to as reference plane), but the IMC is in an inclined orbit. Accordingly, the related NGD model, I30-NGD, I60-NGD, and I120-NGD models, describe the same misalignment between the stars and IMC but without considering the existence of gas disk. 

Compare to the 2-d cases, the sweeping secular resonance of an inclined IMC can still significantly excite disk stars albeit less effective, especially for the highly inclined orbit, e.g., I60 model. Even though there is no gas disk under the coplanar circumstance, the IMC's secular perturbation and mean motion resonances can stimulate some hyper-velocity escapees in the NGD model. However, in the slightly inclined case without gas disk, model I30-NGD, only a few disk stars can be incredibly excited with a periapse distance smaller than $10^{-4}$~pc. As a result, no HVS was ejected. The situation is improved in the I60-NGD model. In the absence of a gaseous disk, the sweeping secular resonance effect does not contribute to any HVS population in this case.  The mutual inclination between the IMC and the disk stars is larger than the threshold value of Lidov-Kozai secular resonance around a point mass potential, which is generally set as $39^{\circ}$. Even though the Galactic bulge potential introduces additional precession, more disk stars are excited and ejected
in the I60-NGD model than the I30-NGD model. 

In Figure \ref{fig:d_i_e}, we focus on the inclination distribution of disk stars under each inclined model after $\sim 6$ Myr evolution. For a clearer contrast, the left panels of Figure \ref{fig:d_i_e} show the results of inclined cases with the gas disk, while the right panels are the corresponding models without the gas disk. As expected, the presence of a gas disk not only introduces 
sweeping secular resonance which leads to efficient $e_\ast$ excitation and ejection of HVSs but also damps the stars' excited 
inclination relative to the disk and preserve their initial $i_\ast$, especially for those residuals at $r_\ast< 0.1$~pc 
(see also in the paper I). 

\begin{figure*}
\epsscale{1}
\plotone{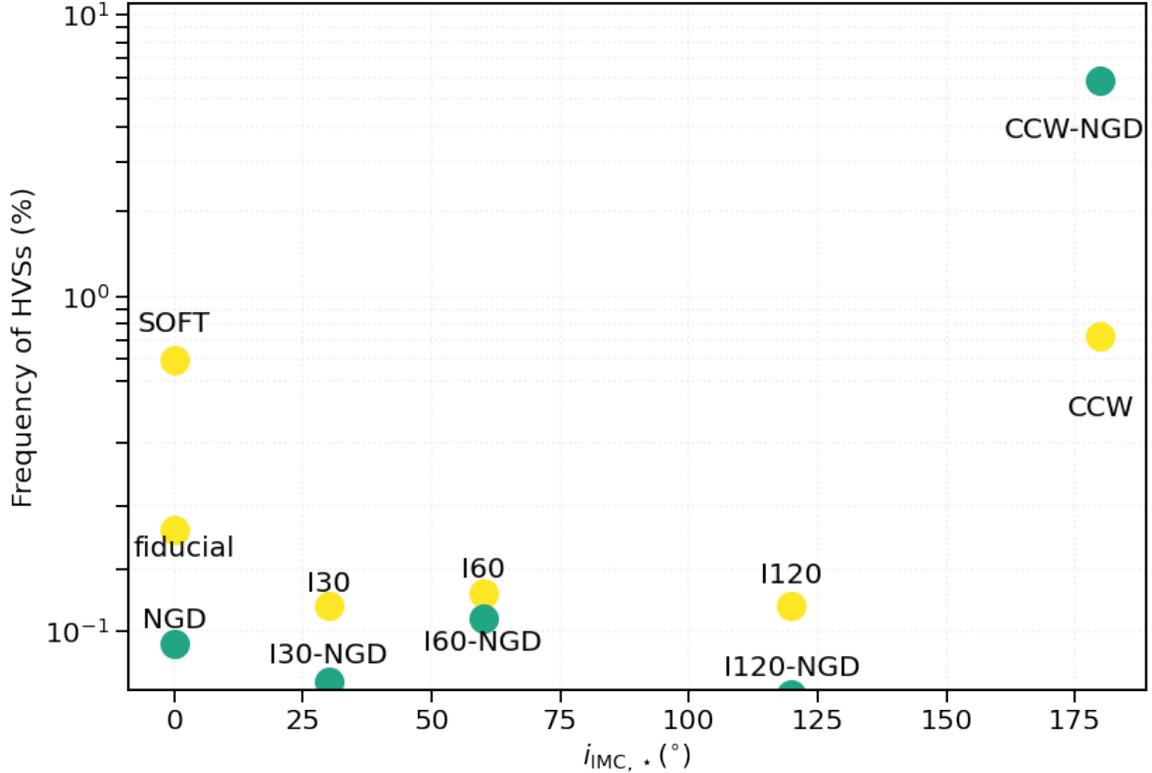}
\caption{The HVSs frequency in each model. The blue dots label the models without gas disk, while the yellow dots represent the models with gas disk. The x-axis, $i_{\rm IMC, \star}$, refers to the initial mutual inclination between the orbits of IMC and disk stars. 
\label{fig:model_frq}}
\end{figure*}

In the paper I, we show that the characteristic timescales for secular evolution for the Galactic center stars depend weakly on 
$a_\ast$. Since they are comparable to the simulated duration ($\sim 6$~Myr) for all the models, typical stars, including those 
with $a_\ast \lesssim 0.1$~pc, went through approximately one to a few secular cycles in both $e_\ast$ and $(i_\ast - i_{\rm IMC})$.  
This pattern is evident in the relatively low $e_\ast$ and the gradient of $(i_\ast-i_{\rm IMC})$ after $\sim 6$ Myr. The range of $e_\ast$ modulation is small for $a_\ast < 0.1$~pc. However, the eccentricity of some more distant stars (with $a_\ast>0.1$~pc) are significantly excited.  These stars also endure large changes in $i_\ast$ from their initial values.  Due to their contribution
to the potential, changes in $i_\ast$ in models with a gas disk are generally smaller than those without gas disk (i.e., the NGD models).



%
\begin{figure*}
\epsscale{1.}
\plotone{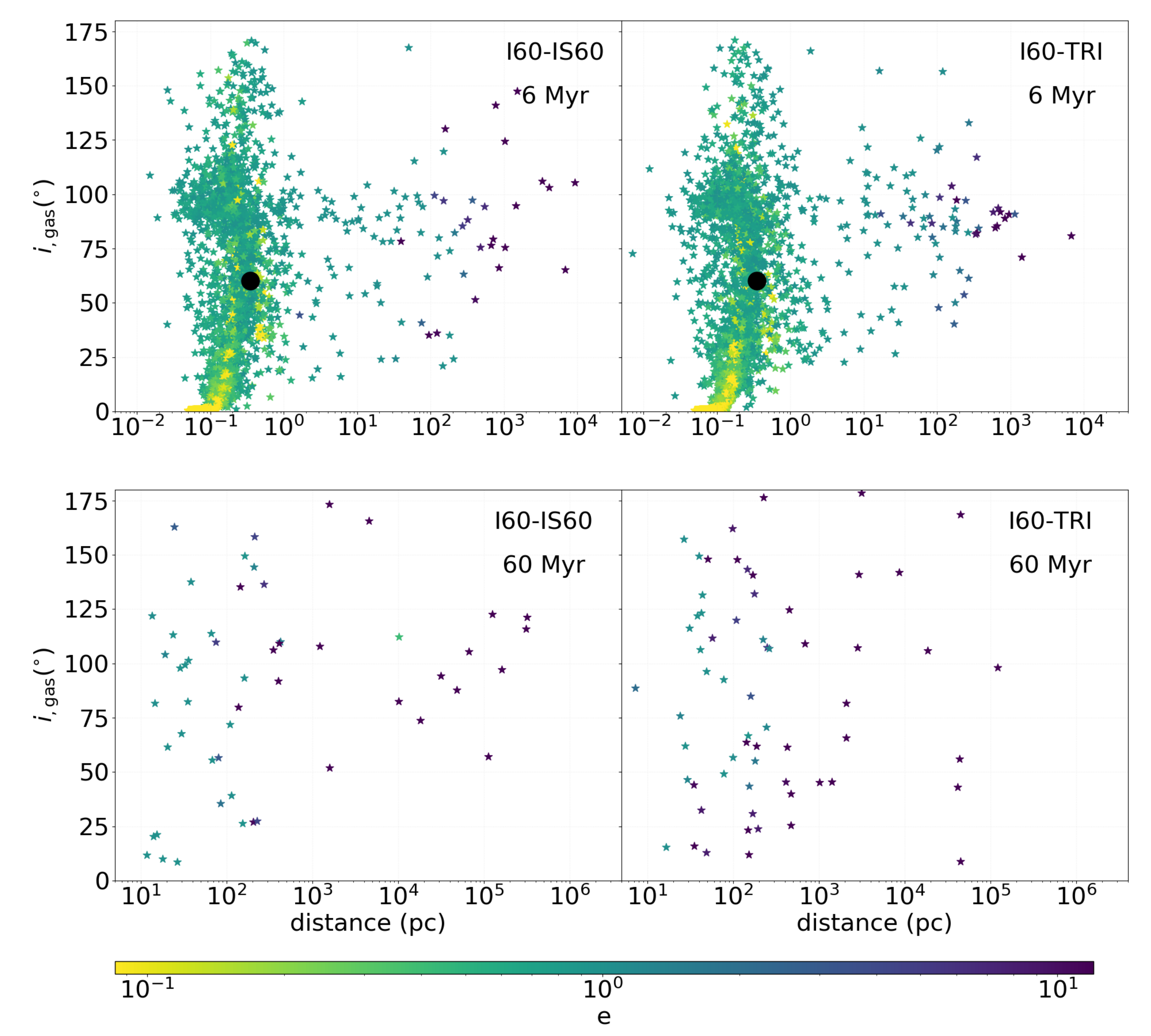}
\caption{The top panels are the same as Figure~\ref{fig:d_i_e}, but with a comparison between the I60-IS60  and the I60-TRI models. $i_{,\rm gas}$ refers to the inclination of the orbits of disk stars (and IMC) relative to the plane of the gas disk, range from $0^{\circ}$ to $180^{\circ}$. The bottom panels keep tracing these escapees' final trajectories in the I60-IS60 and I60-TRI models around 60 Myr.  
\label{fig:d_i_e_tri}}
\end{figure*}
\begin{figure*}
\epsscale{1}
\plotone{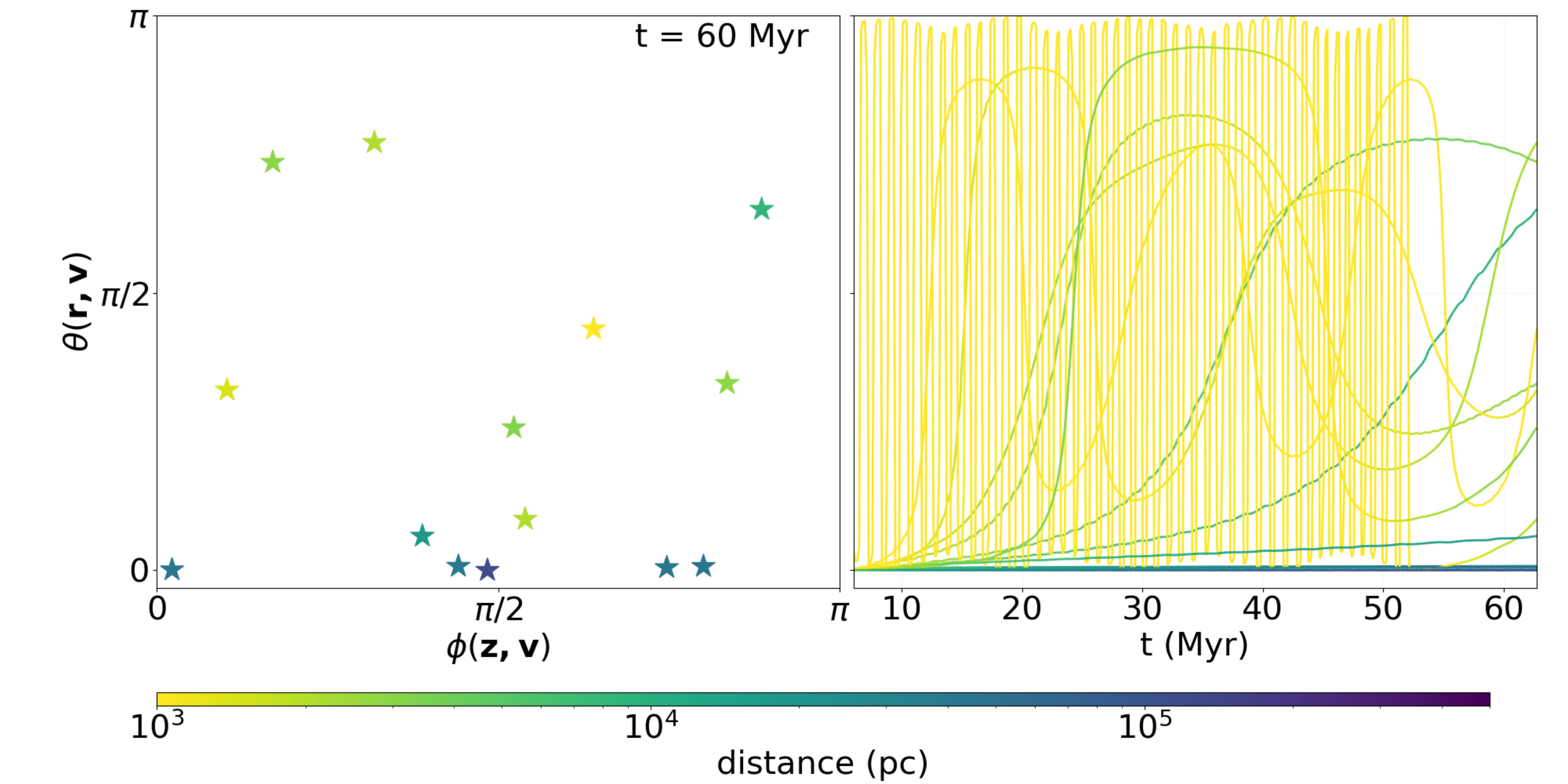}
\caption{{\bf Left panels:} It is the instantaneous trajectories of ejected stars with distance $> 1$ kpc in the I60-TRI model around 60 Myr. The z-axis is the azimuth angle, $\phi$, which is the angle between ${\bf z}$ and ${\bf v}$. The y-axis is the angle $\theta$ shows between the direction of ${\bf r}$ and ${\bf v}$. {\bf Right panels:} The $\theta ({\bf r, v})$ of these ejected stars as a function of time (from $\sim 6$ Myr to $\sim 60$ Myr).
The color indicates the distance from the Galactic center.
\label{fig:the_d_tri}}
\end{figure*}

\begin{figure*}
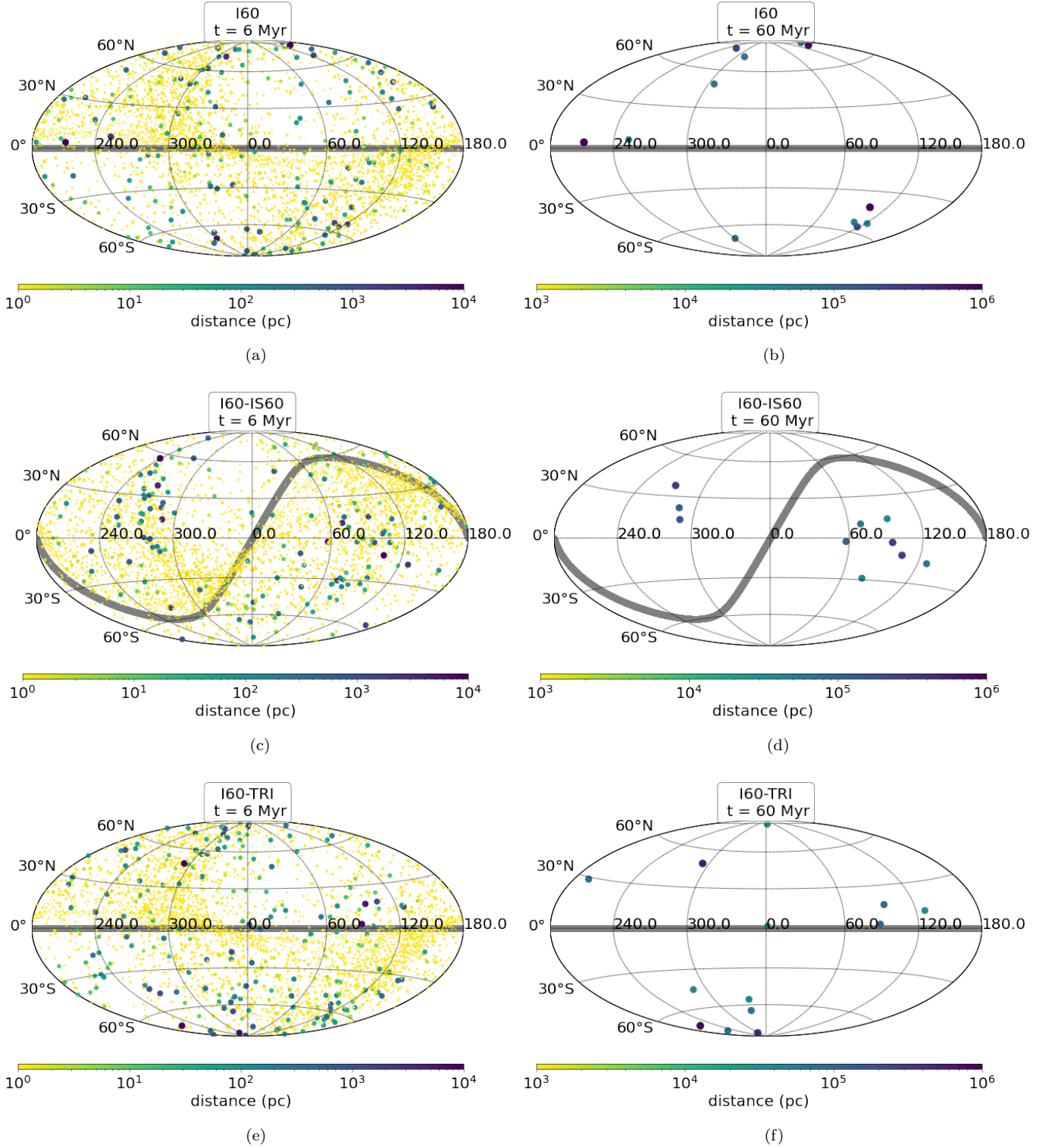

\gridline{\fig{i60-6myr.pdf}{0.5\textwidth}{(a)}
          \fig{i60-60myr.pdf}{0.5\textwidth}{(b)}
}
\gridline{
          \fig{i60-is60-6myr.pdf}{0.5\textwidth}{(c)}
          \fig{i60-is60-60myr.pdf}{0.5\textwidth}{(d)}
}
\gridline{\fig{i60-tri-6myr.pdf}{0.5\textwidth}{(e)}
          \fig{i60-tri-60myr.pdf}{0.5\textwidth}{(f)}         
}
\caption{Spatial distributions of 10000 simulated stars in the I60, I60-IS60, and I60-TRI models at 6 Myr (left panels: a, c, and e) and 60 Myr (right panels: b, d, and f, only stars with the velocity $> 500$ km/s are plotted). All disk stars are projected in the standard Galactic coordinates. The initial orbital plane of disk stars is presented with thick grey line. Their distances to the GC are labeled with colors. For better comparison, those faraway, ejected HVSs are marked with larger sizes, proportional to the logarithm of their distances to the GC. 
\label{fig:gc_projection}}
\end{figure*}

The potential for the gas disk modifies the frequency of both longitudinal and nodal precessions. The disk potential
diminishes on a timescale $\tau_{\rm dep}$ which is also comparable to the duration of the simulation 
and the eccentricity excitation timescale in models I30, I60, and I120. 
Direct comparisons between these models with their diskless counter parts (models I30-NGD, I60-NGD, and I120-NGD)
indicate that changes in both $e_\ast$ and $i_\ast$ are reduced for the close-in stars (with $a_\ast < 0.1$~pc) by 
the disk potential (Figures \ref{fig:a0_rmin_inc} and \ref{fig:d_i_e}).  
Outside $\sim 0.1$~pc, changes in the disk potential during its depletion significantly enlarges the radial extent of 
IMC's secular resonance (cf. equivalent models without the evolving gas disks).  Significant eccentricity excitation 
(leading to $r_{\rm min} < 10^{-3}$~pc) and inclination changes occur for stars with $a_0 \sim 0.15-0.22$~pc, similar 
to the fiducial and SOFT models. This region is populated by stars with a wide range of inclination and relatively 
high eccentricity. This result may be linked to the observed inclination diversity for the young stars around \sgra~\citep{paumard2006, lu2006, lu2009}.  While the eccentricity of many stars is excited to values comparable to that 
observed among the CCWs, few stars acquire $r_{\rm min}$ sufficiently small $(\lesssim 10^{-4}~\rm pc)$ to undergo 
close encounter with the SMBH and then be 
ejected as HVSs.  Although the frequency of HVSs is increased by the presence of the disk, it is also reduced by IMC's
inclined orbit (Figure \ref{fig:model_frq}).

\subsection{HVSs' trajectory in the Galactic halo}
At a distance greater than $\sim 1$~kpc, the HVSs' orbital trajectories are under the influence of the Galactic disk and halo. In all the above inclined models, we assume the initial orbits of all the stars (including those eventually ejected as HVS's) are in the same plane as the Galactic disk. 
We carried out another model in which we
consider the possibility that the stellar disk is inclined $60^{\circ}$ to the Galactic disk \citep{paumard2006} in Figures~\ref{fig:d_i_e_tri} and \ref{fig:gc_projection}. It shows that HVSs' asymptotic trajectories are indeed randomly distributed rather than cluster around the Galactic disk's pole. 

In all but one of the simulations, we adopt a spherically symmetric potential for the Galactic halo. Tri-axial mass distribution in the
Galactic halo \citep{Wegg2019} can lead to the randomization of HVSs' orbits relative to Galactic plane and the reorientation of their
trajectories from radially outward direction relative to the Galactic center \citep{YuMadau2007MNRAS, Brown2015, Du2019}.   
In Figures~\ref{fig:d_i_e_tri}, we plot the stars' orbital inclination relative to their natal disk and their Galactico-centric distance. Their position in the Galactic longitude and latitude, viewed from the Galactic Center is show in Figure \ref{fig:gc_projection}, in a similar way as \cite{Lu2010}.

In model I60-TRI, we use Equation \ref{eq:triaxial} to approximate a tri-axial mass distribution and potential. 
We plot the distribution of the angles 
\begin{equation}
\theta({\bf r, v}) = {\rm cos}^{-1} \left( { {\bf r \cdot v} \over \vert r \vert \vert v \vert} \right)
\ \ \ \ \ \ {\rm and} \ \ \ \ \ \ 
\phi({\bf z, v}) = {\rm cos}^{-1} \left( { {\bf z \cdot v} \over \vert z \vert \vert v \vert} \right)
\end{equation}
distribution for the HVSs after 60 Myr (left panel of Figure~\ref{fig:the_d_tri}).  This sample of 14 HVSs is chosen with 
$r>10^2$ pc and $v>500$ km s$^{-1}$ as in Figure \ref{fig:a0_rmin}.  Six HVSs which are escaping the Galactic 
halo reach $30-60$ kpc with relatively small  $\theta({\bf r, v})$, i.e. their trajectories have not deviated
significantly from the radial direction. Seven HVSs's distance remain within $\sim 10$ kpc.  Their $\theta({\bf r, v})$
are widely distributed as shown by \citet{YuMadau2007MNRAS}.  Although the tri-axial potential has randomized 
their trajectory, there is no obvious indication of clustering along the minor axis of the halo potential.

We also plot the evolution of $\theta ({\bf r, v})$ (right panel of Figure \ref{fig:the_d_tri}) for the same 
HVS sample. The most distant stars are also the fastest moving stars.  Their
trajectories have not deviated significantly from radial motion away from the Galactic center.  However, the orientation
of the closer HVS appears to evolve under the influence of non-spherical halo potential.  Thus, we do not expect the observed proper motion of all the HVSs to be directed away from the Galactic center.

We also find that in the CCW model (not plotted here), most of the escaping stars retain their initial retrograde orbits (with $i_\star = 180^{\circ}$) though the magnitude of their
$e_\star$ is extensively excited. However, some escaping disk stars flip their orbital angular
momentum vector to follow the orbital direction of the IMC.  Since both the disk stars and the IMC
are in the same orbital plane in the CCW model, no star attains an inclined orbit.  Nevertheless,
misalignment between the orbit of the disk stars with respect to the Galactic plane and a tri-axial mass distribution in the halo can also change the spatial distribution of HVS's in this model.

\section{Summary \& Discussion}
\label{sec:summary}

The main motivation of this investigation is to explore the origin of a population of hyper-velocity B stars (HVSs).
These stars appear to be escaping from the Galaxy with velocity $> 500$ km/s.  Their special distribution in the Galaxy
is widely spread out.  The similarities of their spectral class and life expectancy ($\sim 10^7$ yr) with those
of the S stars and disk stars near \sgra~ naturally raise the possibility that they may have been ejected from
the proximity of the SMBH at the center of the Milky Way.

The central conundrum is the mechanisms that enabled the HVSs to gain such enormous kinetic energy and to venture out of the deep gravitational potential imposed by the SMBH.  A  widely accepted scenario proposed by \citet{Hills1988} suggests that an infusion of close binary stars to the proximity of an SMBH can lead to their tidal disruption and the ejection of one component with hyper-velocity.  The main drawback of this scenario in the
present context of the Galactic HVSs is the long timescales required for the diffusion of binary stars through
stellar relaxation processes \citep{lightman1977, Rauch1996, merritt2013}.
 
In the paper I of this series, we show a promising mechanism for rapid eccentricity excitation near \sgra.  We attribute
the main cause to be secular perturbation from an intermediate-mass companion to the super-massive black hole. We speculate the IMC culprit to be IRS 13E.  We show the robustness of this effect for a wide range of IMC's orbital configuration especially if we consider the evolution of the gravitational potential during the
depletion of their natal disk.  We provide analytic calculation and numerical simulation to show that recently formed
(several Myr ago) stars can indeed attain the large observed eccentricities. We showed that the IMC's secular perturbation is over a considerable distance and extensive time such that it can be either a compact stellar cluster or an IMBH with a mass $\sim 10^4 M_\odot$.

In this paper, we extend the IMC scenario for the HVSs' past dynamical evolution. With numerical simulations, we show that the IMC's secular perturbation is effectively exciting the eccentricity of some young stars to near
unity.  Through this action at a distance, the orbital energy of these nearly parabolic stars essentially reduces the minute fraction of their gravitational potential energy at their periastron.  Some stars' periastron approaches to (but do not cross) their tidal disruption radius while their apoastron crosses the IMC's orbit.

Due to the combined perturbation by the IMC and the Galactic bulge, the stars' eccentric orbits precess.  Some stars undergo close encounters with the IMC and endure impulsive 3-body interaction which modifies their trajectory, especially
the impact parameter, $1-e_\ast$, and energy of their orbital energy relative to the SMBH.  In the next close encounter with the SMBH, some stars are slingshot to escape IMC's gravitational confinement as HVSs.   

We examine the robustness of this process for a range of IMC's orbital configurations. The ejection probability ranges
from a few to a fraction of a percent depending on whether the orbits of the IMC and the disk stars are in the same plane
(in either a counter or co-orbiting configuration) or inclined planes.  Although the presence and depletion of a natal disk
promote the ejection of the HVSs, they are not essential.  The asymptotic trajectories of the HVSs are widely distributed
rather than concentrated in a plane over the sky.  If the halo potential is tri-axial, the instantaneous HVSs' direction of motion need not be pointed away from the \sgra, albeit the fastest moving HVSs may follow such pathways.

This IMC-secular-perturbation mechanism differs from the binary-break-up mechanism suggested by Hills since all the stars in
our models are single.  Nevertheless, three body interaction (between the star, IMC, and 
SMBH) plays a pivotal role in the liberation of the HVSs \citep{Yu2003}.  In this work, we
include an additional ingredient of IMC's secular perturbation and sweeping secular resonance as the
main mechanism for rapidly supplying recently formed stars to the neighborhood of the SMBH. 
The main advantage of this scenario for the origin of HVSs is that it can naturally account 
for the youthfulness of the HVSs.

Since the periastron distance of the escaping stars are generally well
outside the tidal disruption radius (Figures \ref{fig:a0_rmin} and \ref{fig:a0_rmin_ccw}),
some young close binary stars may survive their close encounter with the SMBH as they are
ejected to become HVSs.  The detection of
any such binary HVSs can provide a discriminative test since they are excluded by the Hill's 
mechanism \citep{Lu2007}.  

Finally, the results presented here are robust for a wide range of IMC's kinematic properties.  
Similar companions may be common in other galaxies, especially if galaxies evolved through mergers 
(a basic concept of $\Lambda$CDM scenario) and central massive black holes in galaxies are ubiquitous. 
These massive black holes undergo orbital decay toward the center of their host galaxies under the action
of dynamical friction and gas drag. During their approach toward coalescence, similar dynamical effects 
between migrating massive black holes may excite the eccentricity of their surrounding stars
as well as stars along their paths.  The subsequent rapid injection of stars to the proximity 
of the massive black holes can significantly modify the rate of tidal disruption events, the
formation and coalescence of a binary star and stellar-mass black holes (as sources of gravitational 
wave events), and the ejection of hyper-velocity stars and stellar-mass black holes.  We end 
with speculation that such processes may even lead to the possibility of freely floating 
intermediate-mass black holes outside any host galaxies.  The follow-up investigation of these 
related issues will be presented elsewhere.

\acknowledgments
Simulations in this paper made use of the REBOUND code which is publicly available 
at \href{http://github.com/hannorein/rebound}{REBOUND} website.  We thank Tuan Do,
Stefan Gillessen, Mark Morris, and Hangci Du for useful conversation.  
This work is partly supported by the National Key Research and Development Program of China (No. 2018YFA0404501 to SM), by the National Science Foundation of China (Grant No. 11821303, 11761131004 and 11761141012 to SM) and by the China Postdoctoral Science Foundation (Grant No. 2017M610865 to XCZ).
\vspace{5mm}

\software{REBOUND \citep{reinliu2012, rein2019}}

\bibliography{zxc.bib}{}
\bibliographystyle{aasjournal}

\end{CJK*}
\end{document}